\documentclass[iop]{emulateapj}
\usepackage{natbib}
\usepackage{amsmath}

\usepackage{color}
\usepackage[usenames,dvipsnames,svgnames,table]{xcolor}
\usepackage[colorlinks=true,citecolor=blue,linkcolor=blue]{hyperref}

\definecolor{darkblue}{rgb}{0,0,0.9}

\begin{document}

\newcommand{\Aphot}{$\rm{A_{phot}}$} 
\newcommand{\w}{$w$}
\newcommand{\csb}{$\rm{c_{SB}}$}

\newcommand{\kpc}{\mathrm{kpc}}
\newcommand{\keV}{\mathrm{keV}}
\newcommand{\Flux}{\mathrm{Flux}}
\newcommand{\new}{\mathrm{new}}
\newcommand{\old}{\mathrm{old}}
\providecommand{\e}[1]{\ensuremath{\times 10^{#1}}}

\newcommand{\PR}{$P_3/P_0$ }
\def\bfc{\begin{figure*}[ht]\begin{center}}
\def\efc{\end{center}\end{figure*}}

\title{Testing for X-ray--SZ differences and redshift evolution \\in the X-ray morphology of galaxy clusters }

\altaffiltext{\Harvard}{Department of Physics, Harvard University, 17 Oxford Street, Cambridge, MA 02138}
\altaffiltext{\MIT}{MIT Kavli Institute for Astrophysics and Space Research, Massachusetts Institute of Technology, 77 Massachusetts Avenue, Cambridge, MA 02139}
\altaffiltext{\FNAL}{Fermi National Accelerator Laboratory, Batavia, IL 60510-0500, USA}
\altaffiltext{\AAUChicago}{Department of Astronomy and Astrophysics, University of Chicago, Chicago, IL, USA 60637}
\altaffiltext{\KICP}{Kavli Institute for Cosmological Physics, University of Chicago, Chicago, IL, USA 60637}
\altaffiltext{\ANL}{Argonne National Laboratory, 9700 S. Cass Avenue, Argonne, IL, USA 60439}
\altaffiltext{\CfA}{Harvard-Smithsonian Center for Astrophysics, 60 Garden Street, Cambridge, MA 02138}
\altaffiltext{\Huntingdon}{Huntingdon Institute for X-ray Astronomy, LLC}
\altaffiltext{\Munich}{Faculty of Physics, Ludwig-Maximilians-Universit\"{a}t, Scheinerstr.\ 1, 81679 Munich, Germany}
\altaffiltext{\ExcellenceCluster}{Excellence Cluster Universe, Boltzmannstr.\ 2, 85748 Garching, Germany}
\altaffiltext{\MPE}{Max Planck Institute for Extraterrestrial Physics, Giessenbachstr.\ 85748 Garching, Germany}
\altaffiltext{\UMontreal}{D\'{e}partement de Physique, Universit\'{e} de Montr\'{e}al, C.P. 6128, Succ. Centre-Ville, Montr\'{e}al, Qu\'{e}bec H3C 3J7, Canada}
\altaffiltext{\Yale}{Department of Physics, Yale University, New Haven, CT 06520, USA}
\altaffiltext{\YCAA}{Yale Center for Astronomy and Astrophysics, Yale University, New Haven, CT 06520,USA}
\altaffiltext{\Boulder}{Center for Astrophysics and Space Astronomy, Department of Astrophysical and Planetary Science, University of Colorado, Boulder, C0 80309, USA}
\altaffiltext{\NASAAmes}{NASA Ames Research Center, Moffett Field, CA 94035, USA}

\def\Harvard{1}
\def\MIT{2}

\def\FNAL{3}
\def\AAUChicago{4}
\def\KICP{5}
\def\ANL{6}
\def\CfA{7}
\def\Huntingdon{8}
\def\Munich{9}
\def\ExcellenceCluster{10}
\def\MPE{11}
\def\UMontreal{12}
\def\Yale{13}
\def\YCAA{14}
\def\Boulder{15}
\def\NASAAmes{16}

\author{
D.~Nurgaliev\altaffilmark{\Harvard},    
M.~McDonald\altaffilmark{\MIT},
B.~A.~Benson\altaffilmark{\FNAL,\AAUChicago,\KICP},
L.~Bleem\altaffilmark{\KICP,\ANL},
S.~Bocquet\altaffilmark{\KICP,\ANL},
W.~R.~Forman\altaffilmark{\CfA},
G.~P.~Garmire\altaffilmark{\Huntingdon},
N.~Gupta\altaffilmark{\Munich,\ExcellenceCluster,\MPE},
J.~Hlavacek-Larrondo\altaffilmark{\UMontreal},
J.~J.~Mohr\altaffilmark{\Munich,\ExcellenceCluster,\MPE},
D.~Nagai\altaffilmark{\Yale,\YCAA},
D.~Rapetti\altaffilmark{\Boulder,\NASAAmes},
A.~A.~Stark\altaffilmark{\CfA}, 
C.~W.~Stubbs\altaffilmark{\Harvard,\CfA},
and A.~Vikhlinin\altaffilmark{\CfA}
}

\email{nurgaliev@physics.harvard.edu}
\begin{abstract}

We present a quantitative study of the X-ray morphology of galaxy clusters, as a function of their detection method and redshift.  We analyze two separate samples of galaxy clusters: a sample of 36 clusters at $0.35 < z < 0.9$ selected in the X-ray with the \emph{ROSAT} PSPC 400 deg$^2$ survey, and a sample of 90 clusters at $0.25 < z < 1.2$ selected via the Sunyaev-Zel'dovich (SZ) effect with the South Pole Telescope.  
Clusters from both samples have similar-quality \emph{Chandra} observations, which allow us to  
quantify their X-ray morphologies via two distinct methods: centroid shifts ($w$) and photon asymmetry (\Aphot). The latter technique provides nearly unbiased morphology estimates for clusters spanning  a broad range of redshift and data quality. We further compare the X-ray
morphologies of X-ray- and SZ-selected clusters with those of simulated clusters.  We do not find a statistically significant difference in the measured X-ray morphology of X-ray and SZ-selected clusters over the redshift range probed by these samples, suggesting that the two are probing similar populations of clusters. We find that the X-ray morphologies of simulated clusters are statistically indistinguishable from those of X-ray- or SZ-selected clusters, implying that the most important physics for dictating the large-scale gas morphology (outside of the core) is well-approximated in these simulations. Finally, we find no statistically significant redshift evolution in the X-ray morphology (both for observed and simulated clusters), over the range $z \sim 0.3$ to $z \sim 1$, seemingly in contradiction with the redshift-dependent halo merger rate predicted by simulations.

\end{abstract}

\section{Introduction}
\label{sec:cmpcat_intro}

Large scale galaxy cluster surveys can make an important contribution to 
understanding the growth of structure in the Universe, delivering precise 
constraints on the nature of dark matter and dark energy, and providing 
insights into astrophysical processes in clusters. 
The primary interest in 
studying galaxy clusters from the cosmological point of view is in measuring 
their abundance as a function of mass and redshift, although alternative approaches to cluster-based cosmology that do not rely on precise masses have recently been proposed \citep{caldwell16,ntampaka16,pierre16}. The abundance of galaxy 
clusters as a function of mass and redshift currently provides constraints on 
cosmological models and parameters, most importantly matter density $\Omega_{\rm M}$ 
and the normalization of the matter power spectrum $\sigma_8$ \citep[see review by][]{allen11}.  
There are many subtleties, however, in interpreting abundance information from 
cluster surveys.  First, since measuring the total mass is observationally expensive and highly uncertain (i.e., via weak lensing), most studies
use scaling relations to link a low-scatter mass proxy (such as X-ray spectroscopic 
temperature or integrated SZ signal) to its mass.  Both scatter and potential 
biases in the given scaling relation have an effect on constraints when fitting 
cosmological models.  Second, one needs to understand the survey's completeness 
and purity.  Finally, more subtle selection effects such as increased 
sensitivity to a particular sub-class of clusters may play a role.  An example 
of such a bias would be an increased sensitivity of X-ray flux-limited samples 
to cool core clusters: as cool core systems have higher X-ray luminosity than 
non-cool core systems of the same mass, a flux-limited sample can potentially be 
biased towards cool core clusters \citep{hudson10, mittal11, eckert11}.  
The ratio of cool core to non-cool core systems at different redshifts is currently a subject of active research \citep{vikhlinin06b, santos10, samuele11, mcdonald11c, semler12, mcdonald13b}, so this bias effectively 
limits our understanding of completeness in the X-ray flux-limited samples.

Many of the biases implicit in the selection of various galaxy cluster samples are well understood. 
X-ray luminosity is proportional to gas density squared, so X-ray detection is biased towards cool 
core systems that have high central densities. In contrast, the majority of the SZ signal 
originates from outside of the core. Consequently, SZ detection is biased 
towards the large-scale gas properties in the cluster. Because both X-ray and SZ 
detection methods are based on the physical properties of the ICM, they may have 
some common biases \citep{Maughan12,Angulo12,Lin15} that are completely different from the 
optical detection methods which are sensitive to a different component of galaxy 
clusters (i.e., the galaxies). The finer details of each detection method's sensitivity to specific 
cluster morphology or dynamical state are not well understood.

It has been suggested that SZ-selected clusters are more often ``morphologically disturbed'' (i.e., ongoing mergers) than their X-ray-selected counterparts. This line of reasoning stems from 1) the presence of spectacular mergers among the first discovered by the SZ effect, such as El Gordo 
\citep{menanteau11} and PLCKG214.6+37.0
\citep{planck13c}; and 2) an extensive discussion of newly discovered 
mergers in the papers which originated from the \emph{Planck} \emph{XMM-Newton} 
follow-up program \citep{planckxmm11,planckxmm12,planckxmm13}. The 
latter program targeted 51 cluster candidates and led to the confirmation of 43 
candidates, 2 of them being triple systems, and 4 double systems. The 37 
remaining objects had 1) lower X-ray luminosity than expected from scaling 
relations and 2) shallower density profiles than the mean density profiles of 
X-ray detected clusters. These two observations served as the main arguments for 
\emph{Planck's} increased sensitivity to mergers. Indeed, recent studies have shown that \emph{Planck} clusters are, on average, more morphologically disturbed \citep{rossetti16} and have a lower occurrence rate of cool cores \citep{jones15} when compared to X-ray-selected clusters.

While systems discovered by \emph{Planck} do have interesting morphological properties, 
the aforementioned findings do not necessarily indicate an inherent sensitivity of the SZ 
effect to merging clusters. 
The lower central density and luminosity of clusters may be related to greater than 
previously thought intrinsic scatter in these parameters, or factors other than 
merging processes. Several of the double and triple systems discovered by \emph{Planck} 
are clusters overlapping in projection, rather than interacting systems 
(although they still belong to the same supercluster structure). The increased 
sensitivity of \emph{Planck} to such multiple systems is unsurprising due to its 
large beam size and consequent inability to resolve multiple systems. 

Another question that has been extensively discussed in the literature 
is whether cluster morphology depends on redshift. The motivation for these 
studies is the connection of merger rate (and consequently morphology) to the mean
matter density $\Omega_{\rm M}$. The morphology-cosmology connection that was 
analytically developed by \citet{richstone92} and then confirmed in 
simulations by \cite{evrard93} and \cite{jing95} predicted that clusters in low 
$\Omega_{\rm M}$ models are much more regular and spherically symmetric than those in 
$\Omega_{\rm M}=1$ models. Consequently, there were efforts to constrain $\Omega_{\rm M}$ by 
finding the fraction of clusters with significant level of substructure as 
defined by various substructure statistics: \citet{mohr95} used centroid 
shifts, \citet{buote95} used power ratios, and \citet{schuecker01} used a trio of tests which quantify mirror symmetry, azimuthal symmetry, and radial elongation. 
This approach to constraining cosmological parameters through substructure rates has not been as successful as other cluster-based cosmology tests owing to difficulties in robustly defining ``significant levels of substructure'', connecting observable substructure measures to theoretical merger definitions \citep{buote95}, and insufficiently low numbers of observed and simulated clusters for these tests. 
%

Modern halo abundance measurements provide much more precise constraints on 
$\Omega_{\rm M}$ than those obtained by merger fraction studies. Nevertheless, the 
question of substructure evolution in galaxy clusters is still relevant. The majority of 
studies have reported statistically significant evolution in cluster morphology \citep{jeltema05,andersson09,mann12}; while a smaller number \citep[e.g.][]{weissmann13b,mantz15} 
arrived at the conclusion that clusters at low- and high-$z$ are consistent with no morphological 
evolution.  \citet{weissmann13b} performed a study of substructure evolution 
similar to ours, which is described later, using a slightly different cluster sample and substructure 
statistics, but arriving at similar results (See Sec.~\ref{sec:discussion} for 
more details).

Our objectives in this paper are to test for any evidence of a difference in 
dynamical state between X-ray and SZ-selected clusters, low-z and high-z 
clusters, and observed and simulated clusters. The difference between X-ray and 
SZ-selected samples is of particular interest if we wish to combine the X-ray 
and SZ samples in order to obtain better statistics for various studies of 
cluster properties. In \S2, we describe the three cluster samples used in this paper, from the South Pole Telescope, ROSAT PSPC 400 deg$^2$ survey, and from numerical simulations. In \S3 we describe our methodology for quantifying X-ray morphology and the various tests that we will perform. In \S4 we will discuss results of these tests, focusing on the key questions of whether or not X-ray- and SZ-selected clusters are statistically different in terms of their X-ray morphology, whether simulated and real clusters have statistically different morphology, and whether there is any measurable redshift evolution in X-ray morphology. In \S5 we will discuss these results, placing them in context of previous work and considering their implications. We will conclude in \S6 with a brief summary and look forward to future studies.

Throughout this work, we assume a flat $\Lambda$CDM cosmology with H$_0$=70 km s$^{-1}$ Mpc$^{-1}$ and $\Omega_{\rm M}$ = 0.27.

\section{The data}
\label{sec:cmpcat_data}

\subsection{Observations}

The basis of this study is a subsample of 90 galaxy clusters, drawn from the larger sample of 516 galaxy clusters in the 2500 deg$^2$ South Pole Telescope (SPT) survey of \cite{Bleem15}. These 90 clusters, which were amongst the most massive of the SPT-selected clusters, all have uniform-depth \emph{Chandra} observations, as summarized in \cite{mcdonald13b,mcdonald14c}. X-ray observations of these clusters were obtained primarily via a Chandra X-ray Visionary Project (PI B. Benson). Clusters in the SPT sample span the redshift range $0.25 < z < 1.2$ and the mass range $2 \times 10^{14}$ M$_{\Sun} \lesssim$  M$_{500} \lesssim 2 \times 10^{15}$ M$_{\Sun}$, where M$_{500}$ is the total mass within $r_{500}$, and $r_{500}$ is the radius within which the average enclosed density is 500 times the critical density. M$_{500}$ here is derived from the Y$_X$--M relation, following \cite{Andersson11}. The median redshift and M$_{500}$ for this sample are 0.59 and $4.6 \times 10^{14}$ M$_{\Sun}$ respectively.

We use the high-$z$ part of the ROSAT PSPC 400 deg$^2$ cluster survey \citep{Burenin07}, abbreviated hereafter as 400d, for our X-ray-selected sample.  This sample consists of 36 clusters in the redshift range $0.35 < z < 0.9$ and mass range $10^{14}$ M$_{\Sun} <$ M$_{500} < 5 \times 10^{14}$ M$_{\Sun}$. The median redshift and mass of this sample are 0.48 and $2.6  \times 10^{14}$ M$_{\Sun}$ respectively. The masses (M$_{500}$) for these clusters were determined in the same way as for the SPT-selected clusters, using the same pipeline and scaling relations. These X-ray-selected clusters have a distinct lack of strong central density cusps at $z>0.5$ \citep{Vikhlinin07}, similar to what is observed in SPT-selected clusters \citep{mcdonald13b}. 

\begin{figure}[ht]
    \epsscale{1.1}
    \plotone{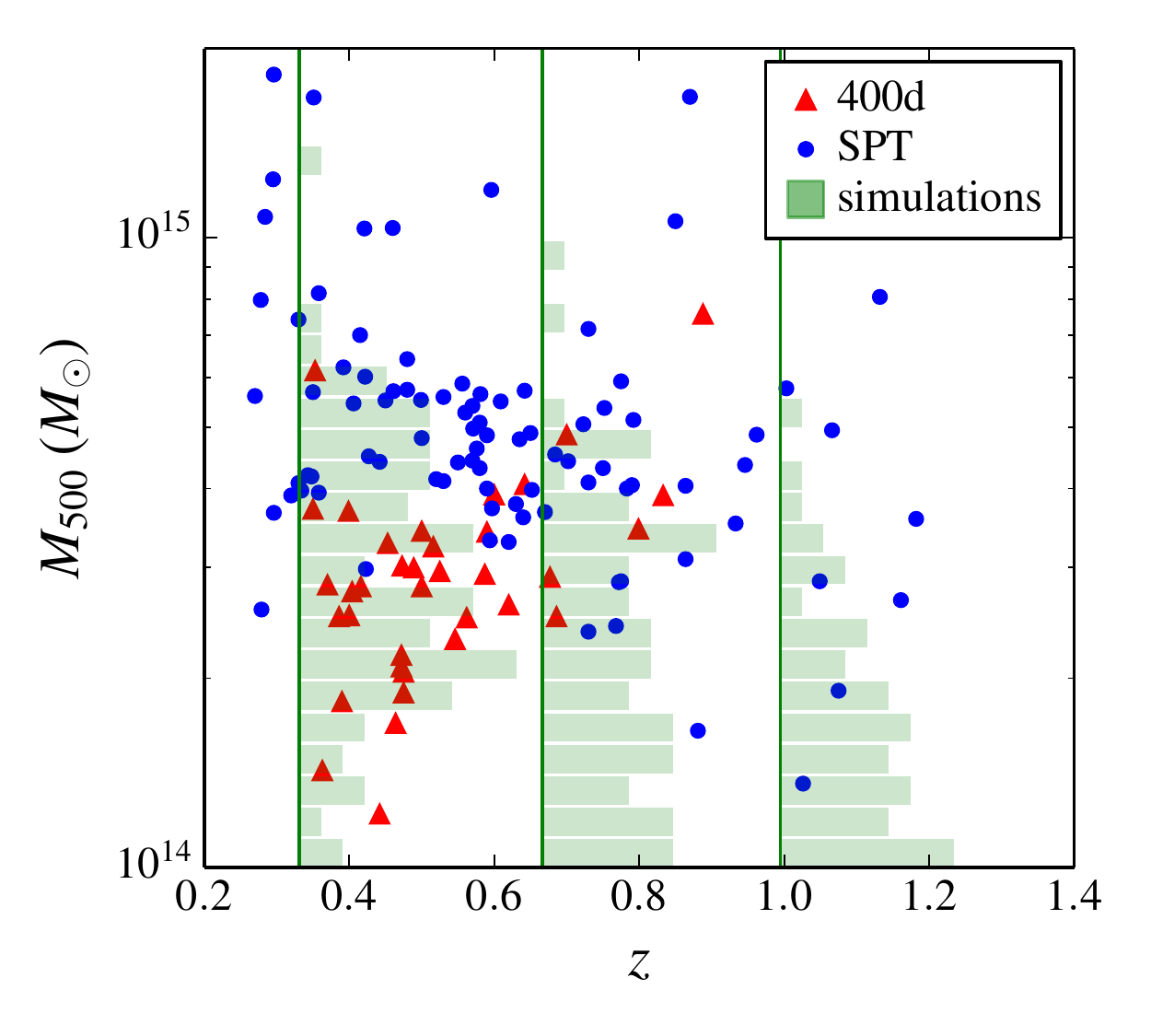}
    \caption{
        Masses and redshifts for SPT, 400d, and simulated samples. 
        Red and blue dots show individual objects in the 400d and SPT samples, respectively.
        Green histograms show the distributions of masses for simulated clusters in 3
        arbitrarily chosen redshifts bins --- there are a total of 85 simulated clusters used for comparison to the data.
        The SPT and 400d samples have a reasonable
        overlap for $z<0.6$; simulated clusters have a reasonable overlap with 
        the observed clusters at all redshifts. We will show in \S4.2 that the slightly different mass range for these samples should not significantly bias the distribution of observed morphologies.
    }
    \label{fig:cmpcat_zm}
\end{figure}

Fig.~\ref{fig:cmpcat_zm} shows the distribution of SPT and 400d clusters in the 
($z$, M$_{500}$) plane. There are fewer clusters at $z>0.6$ in the 400d 
sample, due to the fact that it is flux limited. For a fair comparison (which would be free of redshift 
evolution effects) we will compare morphologies in the $z<0.6$ subsamples for
both catalogs --- we will return to this point later when we define comparison samples. In this low-$z$ subsample, the median masses of the 400d and SPT samples are 2.8 and $5.3 \times 10^{14}$ M$_{\odot}$, respectively.
This figure highlights the overlap in mass and redshift between the two observational samples, along with the simulated clusters that will be described below.

Both samples have similar quality \emph{Chandra} observations. Exposures are typically sufficient to obtain $\sim$1500--2000 X-ray source counts \citep[see][]{mcdonald14c}. The high-resolution \emph{Chandra} imaging with sufficient photon statistics is crucial to detect substructure in galaxy clusters \citep{Nurgaliev13}. The 400d and SPT samples are currently among the best available samples of high-redshift clusters with clear selection criteria and high-quality X-ray follow-up. Significant overlap in the redshift and mass ranges allow us to directly compare clusters in these two samples.

The X-ray data reduction steps for both samples are equivalent to those 
described in \citet{Vikhlinin09a}, \cite{Andersson11}, and \cite{mcdonald13b}. Using \textsc{ciao} v4.7 and \textsc{caldb} v4.7.1, we first filter data for flares, before applying the latest calibration corrections. Point sources are identified using an automated routine following a wavelet decomposition technique \citep{Vikhlinin98}, and then visually inspected. Cluster centers are chosen to be the brightest pixel after convolution with a Gaussian kernel with $\sigma = 40$ kpc, following \cite{Nurgaliev13}.

\subsection{Simulations}

We analyze mock {\em Chandra} observations of massive galaxy clusters extracted 
from the \textit{Omega500} simulation \citep{Nelson14}. Below, we briefly summarize the main elements of the  
\textit{Omega500} simulation and mock {\em Chandra} observations of simulated clusters, and 
refer the readers to \citet{Nelson14} and \citet{Nagai07}, respectively, for 
further details. 

The \textit{Omega500} simulation is a large cosmological hydrodynamic simulation performed with 
the Adaptive Refinement Tree (ART) N-body+hydrodynamics code 
\citep{Kravtsov99, Kravtsov02, Rudd08}. In order to achieve the dynamic
ranges necessary to resolve the cores of massive halos, adaptive refinement in space and 
time and non-adaptive refinement in mass \citep{Klypin01} are used. The 
simulation has a co-moving box length of $500h^{-1}$ Mpc and a maximum co-moving 
spatial resolution of $3.8h^{-ˆ'1}$~kpc, (where $h \equiv$ 0.01H$_0$) and is performed in a flat $\Lambda$CDM 
model with the WMAP five-year cosmological parameters \citep{Komatsu09}. The simulations include gravity, collisionless dynamics of dark matter and stars, gas dynamics, star formation, metal enrichment, SN feedback, advection of metals, metallicity-dependent radiative cooling, and UV heating due to a cosmological ionizing background. Some relevant physical processes, including AGN feedback, magnetic fields, and cosmic rays were not included.

For each cluster with M$_{500} > 10^{14}$ M$_{\odot}$, of which there are 85, the regions within 
$5\times r_{\rm vir}$ are re-simulated with high spatial and mass resolution 
using the multiple mass resolution technique. This leads to a maxiphysical resolution of 5
The resulting simulation has 2048$^3$ spatial elements,
allowing a corresponding dark matter particle mass of $1.09 \times 10^9\, 
h^{-1}$M$_{\odot}$. Herein we analyze the simulated galaxy clusters at expansion factors $a=0.5014, 0.6001, 0.7511, 0.9764$
corresponding to redshifts $z=0.99, 0.66, 0.33, 0.02$. The distribution of masses of simulated clusters at each of these epochs is shown on Fig.\ \ref{fig:cmpcat_zm}.

For each simulated cluster, we analyze mock {\em Chandra} data. We 
first generate X-ray flux maps for each of the simulated clusters. We compute the 
X-ray emissivity, using the MEKAL plasma code 
\citep{Mewe86,Kaastra93,Liedahl95} and the solar abundance table from 
\citet{Anders89}.  We multiply the plasma spectrum by the photoelectric 
absorption corresponding to a hydrogen column density of $N_{\mathrm{H}}=2\times 
10^{20}$~cm$^{-2}$. We then convolve the emission spectrum with the response of 
the {\em Chandra} ACIS-I CCDs and draw photons from each position and spectral 
channel according to a Poisson distribution with exposure time necessary to 
achieve similar total number of counts as in the SPT and 400d samples.
We project X-ray emission from hydrodynamical cells along 3 perpendicular lines  
of sight within three times the virial radius around the cluster center, which will allow us to determine how projection affects the determination of X-ray morphology. For 
$z=0.33, 0.66, 0.99$ the spatial resolution of the simulations correspond to 4.0, 3.3, and 2.7 kpc, which is higher resolution than the \emph{Chandra} PSF at these redshifts. For $z=0.02$, we additionally displaced each 
photon in the map by a Gaussian noise with width 10 kpc, to approximate the spatial resolution we achieve for high-$z$ systems.

We generate X-ray maps 100 times for each cluster, redshift, and projection, 
choosing the exposure times so that the distribution of total number of photons 
within an annulus of $0.15-1 R_{500}$ mimics the corresponding distribution of observed counts in the SPT and 
400d samples \citep[see][]{mcdonald14c}. 
This effectively simulates all possible variations in observation quality that are present in the 400d and SPT samples and minimizes the effect of observational S/N on the distribution of the substructure statistics.  
For the purpose of substructure comparison, we treat each simulated X-ray map as an independent object, therefore 
having 85 (clusters) $\times$ 3 (projections) $\times$ 100 (iterations) $=$ 2550 values of the substructure statistic for each of the 4 epochs.

\section{Methods} 
\label{sec:methods}

\subsection{Measuring \Aphot\ and $w$ for Data}
In this analysis, we utilize two morphological parameters that trace the 
degree of cluster disturbance: photon asymmetry (\Aphot; \citet{Nurgaliev13})
and centroid shifts (\w; \citet{Mohr93}). 
Briefly, \Aphot\ quantifies the amount of asymmetry by comparing the cumulative photon count distribution as a function of azimuth for a given annulus to a uniform distribution, computing a probability that these two distributions are different for each annulus using the nonparametric Watson test \citep[for a complete description of this test, see][]{feigelson12, pewsey15}.
%
%
On the other hand, \w\ is a measure of how much the X-ray centroid moves over some radial range (see \citet{Mohr93} for a more detailed description).
\Aphot\ and \w\ show a significant 
degree of correlation with each other, and also with by-eye classification of cluster morphology \citep{Nurgaliev13}. Both are sensitive to spatial irregularities in X-ray 
emission in the plane of the sky. By design, \Aphot\ has more statistical power in resolving 
substructure and is able to produce more consistent results independent of the 
quality of observation (such as exposure, background level, etc). On the other hand, \w\ is more well established in the literature as a widely-used substructure statistic, so we include it in our tests for comparison to these other works.

In Tables \ref{table1} and \ref{table2} we provide asymmetry measurements derived from \Aphot\ and \w\ for each galaxy cluster in the 400d (Table \ref{table1}) and SPT (Table \ref{table2}) samples. These results will be used for the remainder of the study, and are provided here to aid in future studies that wish to isolate a relaxed/disturbed subsample of galaxy clusters over a large range in redshift.

\subsection{Calibrating \Aphot\ With Simulated Clusters}

\begin{figure}[b]
    \epsscale{1.1}
    \plotone{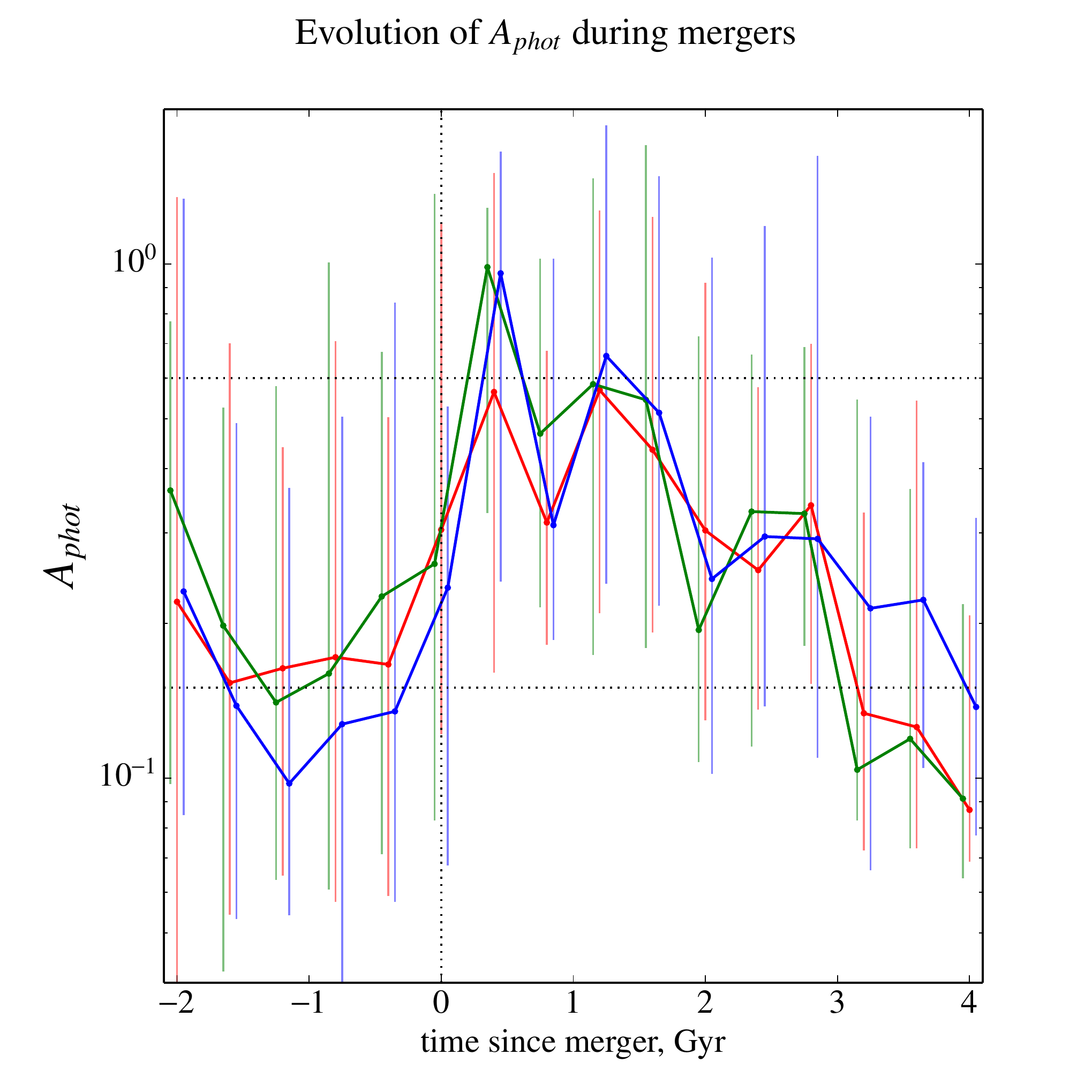}
    \caption{
        Evolution of \Aphot\ for 26 simulated clusters over the course of a major ($M_1/M_2 > 0.5$) merger. The solid colored lines shows the median value of \Aphot\ as a function of time since merger for all 26 clusters along three different sight lines (blue and green are in the plane of the sky, while red is along the line of sight), while the vertical bars show the 1$\sigma$ scatter. Horizontal lines at 0.2 and 0.6 show our divisions for relaxed and disturbed clusters, respectively. The vertical dotted line corresponds to the start of the merger. We find that, within 1\,Gyr of the merger, $\sim$70\% of the simulated images (over all three projections) have \Aphot\ $>$ 0.6, while $<$15\% of the pre- or post-merger systems have such high values, making \Aphot\ $>$ 0.6 a reasonable proxy for ongoing mergers.}
    \label{fig:merger}
\end{figure}

To measure how well photon asymmetry, \Aphot, correlates with the dynamical state of a cluster, and to determine realistic cutoff values for ``disturbed'' and ``relaxed'' systems, we consider a set of simulated major mergers (M$_1$/M$_2 > 0.5$) of massive clusters from the \textit{Omega500} simulations \citep{Kravtsov99, Kravtsov02, Rudd08}. For each of 26 mergers, we produce X-ray photon maps along 3 different projections and at 16 different times, starting $\sim$2 Gyr before the first core passage and ending $\sim$5 Gyr after. At each time step, we compute \Aphot\ from the simulated observations in the same way that we do for the data. Fig.~\ref{fig:merger} shows the results of this study.
For all 78 mergers (26 clusters $\times$ 3 projections), the measured \Aphot\ increases dramatically immediately following the merger, with the median cluster having \Aphot\ $\sim$ 0.9 immediately after the merger and \Aphot\ $>$ 0.6 for $\sim$1 Gyr after. We choose \Aphot\ $=$ 0.6 as a reasonable threshold for a ``disturbed'' cluster, due to the fact that $\sim$70\% (54/78) of our simulated systems (combining all three projections) cross this threshold immediately following a major merger -- we do not expect to identify 100\% of merging systems with a 2-D metric such as \Aphot\ due to the fact that a significant fraction of systems will be merging along the line of sight. We note that the median value of \Aphot\ immediately after a line-of-sight merger is $\sim$0.6.

Before and after the merger, the median value of \Aphot\ fluctuates about $\sim$0.2, indicating that this may provide a reasonable threshold for ``relaxed'', depending on how strict one wants to be with that identifier \citep[see][for a more strict classification]{mantz15}. We note that during this pre- and post-merger phase, the measured \Aphot\ is inconsistent with $>$0.6 for $\gtrsim$85\% of systems, meaning that we expect little contamination in the ``disturbed'' sample from systems that are not currently undergoing major mergers.

Based on these arguments, we infer that an elevated \Aphot\ ($>$0.6) is an adequate proxy of a major merger within the past $\sim$1--2 Gyr.


%
\begin{center}
\begin{longtable}{c c c c c}
\caption[]{X-ray asymmetry measurements for a sample of 36 X-ray-\\selected clusters from the ROSAT PSPC 400 deg$^2$ survey.}
\label{table1}\\
\hline\hline
Name & z & M$_{500}$ & \Aphot\ & \w\ \\
 & & [10$^{14}$ M$_{\odot}$] & & \\
\hline
\endfirsthead
\multicolumn{5}{c}%
{{\textbf{Table \thetable{}} --- continued from previous page}} \\
\hline\hline
Name & z & M$_{500}$ & \Aphot\ & \w\ \\
 & & [10$^{14}$ M$_{\odot}$] & & \\
\hline
\endhead
%
cl0302m0423 & 0.350 & 3.7 & 0.06$^{+0.02}_{-0.03}$ & 0.005$^{+0.001}_{-0.001}$ \\
cl1212p2733 & 0.353 & 6.2 & 0.05$^{+0.03}_{-0.11}$ & 0.018$^{+0.003}_{-0.004}$ \\
cl0350m3801 & 0.363 & 1.4 & 0.14$^{+0.07}_{-0.11}$ & 0.026$^{+0.010}_{-0.008}$ \\
cl0318m0302 & 0.370 & 2.8 & 0.41$^{+0.10}_{-0.09}$ & 0.025$^{+0.004}_{-0.003}$ \\
cl0159p0030 & 0.386 & 2.5 & 0.01$^{+0.03}_{-0.06}$ & 0.010$^{+0.003}_{-0.004}$ \\
cl0958p4702 & 0.390 & 1.8 & 0.05$^{+0.05}_{-0.09}$ & 0.015$^{+0.003}_{-0.005}$ \\
cl0809p2811 & 0.399 & 3.7 & 0.63$^{+0.20}_{-0.24}$ & 0.059$^{+0.008}_{-0.008}$ \\
cl1416p4446 & 0.400 & 2.5 & 0.11$^{+0.02}_{-0.03}$ & 0.014$^{+0.002}_{-0.002}$ \\
cl1312p3900 & 0.404 & 2.8 & 0.22$^{+0.15}_{-0.27}$ & 0.041$^{+0.010}_{-0.011}$ \\
cl1003p3253 & 0.416 & 2.8 & 0.23$^{+0.06}_{-0.08}$ & 0.028$^{+0.003}_{-0.004}$ \\
cl0141m3034 & 0.442 & 1.2 & 0.29$^{+0.18}_{-0.45}$ & 0.091$^{+0.018}_{-0.020}$ \\
cl1701p6414 & 0.453 & 3.3 & 0.14$^{+0.02}_{-0.03}$ & 0.022$^{+0.003}_{-0.002}$ \\
cl1641p4001 & 0.464 & 1.7 & 0.01$^{+0.04}_{-0.06}$ & 0.015$^{+0.004}_{-0.005}$ \\
cl0522m3624 & 0.472 & 2.2 & 0.06$^{+0.05}_{-0.07}$ & 0.013$^{+0.003}_{-0.005}$ \\
cl1222p2709 & 0.472 & 2.1 & 0.04$^{+0.03}_{-0.05}$ & 0.011$^{+0.003}_{-0.004}$ \\
cl0355m3741 & 0.473 & 3.0 & 0.05$^{+0.03}_{-0.04}$ & 0.016$^{+0.004}_{-0.004}$ \\
cl0853p5759 & 0.475 & 2.0 & 0.32$^{+0.10}_{-0.21}$ & 0.051$^{+0.007}_{-0.009}$ \\
cl0333m2456 & 0.475 & 1.9 & 0.25$^{+0.22}_{-0.52}$ & 0.023$^{+0.006}_{-0.010}$ \\
cl0926p1242 & 0.489 & 3.0 & 0.07$^{+0.02}_{-0.03}$ & 0.010$^{+0.003}_{-0.003}$ \\
cl0030p2618 & 0.500 & 3.4 & 0.07$^{+0.08}_{-0.13}$ & 0.013$^{+0.006}_{-0.005}$ \\
cl1002p6858 & 0.500 & 2.8 & 0.07$^{+0.08}_{-0.07}$ & 0.021$^{+0.004}_{-0.006}$ \\
cl1524p0957 & 0.516 & 3.2 & 0.67$^{+0.14}_{-0.13}$ & 0.055$^{+0.007}_{-0.006}$ \\
cl1357p6232 & 0.525 & 3.0 & 0.20$^{+0.06}_{-0.09}$ & 0.022$^{+0.003}_{-0.004}$ \\
cl1354m0221 & 0.546 & 2.3 & 0.26$^{+0.10}_{-0.14}$ & 0.033$^{+0.007}_{-0.006}$ \\
cl1120p2326 & 0.562 & 2.5 & 0.46$^{+0.18}_{-0.36}$ & 0.041$^{+0.007}_{-0.006}$ \\
cl0956p4107 & 0.587 & 2.9 & 0.83$^{+0.18}_{-0.25}$ & 0.042$^{+0.008}_{-0.008}$ \\
cl0328m2140 & 0.590 & 3.4 & 0.08$^{+0.03}_{-0.06}$ & 0.016$^{+0.003}_{-0.003}$ \\
cl1120p4318 & 0.600 & 3.9 & 0.23$^{+0.14}_{-0.20}$ & 0.024$^{+0.005}_{-0.003}$ \\
cl1334p5031 & 0.620 & 2.6 & 0.09$^{+0.18}_{-0.23}$ & 0.016$^{+0.005}_{-0.007}$ \\
cl0542m4100 & 0.642 & 4.1 & 0.53$^{+0.17}_{-0.27}$ & 0.042$^{+0.004}_{-0.005}$ \\
cl1202p5751 & 0.677 & 2.9 & 0.25$^{+0.07}_{-0.07}$ & 0.053$^{+0.008}_{-0.009}$ \\
cl0405m4100 & 0.686 & 2.5 & 0.03$^{+0.03}_{-0.04}$ & 0.008$^{+0.002}_{-0.002}$ \\
cl1221p4918 & 0.700 & 4.9 & 0.14$^{+0.08}_{-0.19}$ & 0.019$^{+0.004}_{-0.005}$ \\
cl0230p1836 & 0.799 & 3.5 & 1.29$^{+0.26}_{-0.22}$ & 0.098$^{+0.007}_{-0.008}$ \\
cl0152m1358 & 0.833 & 3.9 & 2.27$^{+0.22}_{-0.24}$ & 0.155$^{+0.014}_{-0.021}$ \\
cl1226p3332 & 0.888 & 7.6 & 0.10$^{+0.02}_{-0.02}$ & 0.014$^{+0.001}_{-0.001}$ \\
\hline
\end{longtable}       
\end{center}   

\subsection{Statistical Comparisons}
To judge the similarity of cluster subsamples with different selections or redshifts we use both the 2-sample Kolmogorov-Smirnov (KS) and the Anderson-Darling (AD) tests
for the empirical distributions of measured substructure parameters. The values
for these statistics are converted to the p-value of the null hypothesis (that 
these two empirical distributions come from the same underlying distribution).  
Common practice is to reject the null hypothesis for $p < 0.05/N$, where $N$ is the number 
of tests being conducted (the so-called ``Bonferroni correction''). We have tested that using more advanced methods for multiple hypotheses p-value adjustments, such as the Benjamini-Hochberg False Discovery Rate \citep{benjamini95}, does not change any of the conclusions of this paper.

\begin{center}
\begin{longtable}{c c c c c}
\caption[]{X-ray asymmetry measurements for a sample of 90 SZ-selected \\clusters from the South Pole Telescope 2500 deg$^2$ survey.}
\label{table2}\\
\hline\hline
Name & z & M$_{500}$ & \Aphot\ & \w\ \\
 & & [10$^{14}$ M$_{\odot}$] & & \\
\hline
\endfirsthead
\multicolumn{5}{c}%
{{\textbf{Table \thetable{}} --- continued from previous page}} \\
\hline\hline
Name & z & M$_{500}$ & \Aphot\ & \w\ \\
 & & [10$^{14}$ M$_{\odot}$] & & \\
\hline
\endhead
SPT-CLJ0000-5748 & 0.702 & 4.4 & 0.07$^{+0.04}_{-0.05}$ & 0.009$^{+0.002}_{-0.003}$ \\
SPT-CLJ0013-4906 & 0.406 & 5.5 & 0.66$^{+0.06}_{-0.06}$ & 0.029$^{+0.002}_{-0.002}$ \\
SPT-CLJ0014-4952 & 0.752 & 5.4 & 1.44$^{+0.22}_{-0.20}$ & 0.101$^{+0.008}_{-0.008}$ \\
SPT-CLJ0033-6326 & 0.597 & 3.7 & 0.05$^{+0.04}_{-0.06}$ & 0.024$^{+0.006}_{-0.006}$ \\
SPT-CLJ0037-5047 & 1.026 & 1.4 & 0.12$^{+0.23}_{-0.46}$ & 0.066$^{+0.014}_{-0.025}$ \\
SPT-CLJ0040-4407 & 0.350 & 5.7 & 0.06$^{+0.03}_{-0.05}$ & 0.009$^{+0.002}_{-0.003}$ \\
SPT-CLJ0058-6145 & 0.864 & 3.1 & 0.05$^{+0.04}_{-0.07}$ & 0.025$^{+0.005}_{-0.006}$ \\
SPT-CLJ0102-4603 & 0.772 & 2.8 & 0.68$^{+0.18}_{-0.17}$ & 0.074$^{+0.010}_{-0.009}$ \\
SPT-CLJ0102-4915 & 0.870 & 16.8 & 2.46$^{+0.02}_{-0.02}$ & 0.094$^{+0.000}_{-0.000}$ \\
SPT-CLJ0106-5943 & 0.348 & 4.2 & 0.14$^{+0.03}_{-0.04}$ & 0.017$^{+0.003}_{-0.003}$ \\
SPT-CLJ0123-4821 & 0.620 & 3.3 & 0.28$^{+0.09}_{-0.12}$ & 0.039$^{+0.005}_{-0.007}$ \\
SPT-CLJ0142-5032 & 0.730 & 7.2 & 0.33$^{+0.11}_{-0.21}$ & 0.028$^{+0.005}_{-0.006}$ \\
SPT-CLJ0151-5954 & 1.049 & 2.9 & 1.26$^{+0.60}_{-0.33}$ & 0.055$^{+0.012}_{-0.028}$ \\
SPT-CLJ0200-4852 & 0.499 & 5.5 & 0.43$^{+0.13}_{-0.16}$ & 0.033$^{+0.006}_{-0.004}$ \\
SPT-CLJ0212-4657 & 0.640 & 3.6 & 1.11$^{+0.15}_{-0.16}$ & 0.075$^{+0.006}_{-0.005}$ \\
SPT-CLJ0217-5245 & 0.343 & 4.2 & 1.00$^{+0.16}_{-0.17}$ & 0.093$^{+0.007}_{-0.006}$ \\
SPT-CLJ0232-4421 & 0.284 & 10.8 & 0.33$^{+0.02}_{-0.02}$ & 0.017$^{+0.001}_{-0.001}$ \\
SPT-CLJ0232-5257 & 0.556 & 5.9 & 0.30$^{+0.06}_{-0.12}$ & 0.043$^{+0.006}_{-0.006}$ \\
SPT-CLJ0234-5831 & 0.415 & 7.0 & 0.15$^{+0.04}_{-0.06}$ & 0.007$^{+0.002}_{-0.002}$ \\
SPT-CLJ0235-5121 & 0.278 & 8.0 & 0.22$^{+0.04}_{-0.05}$ & 0.031$^{+0.003}_{-0.003}$ \\
SPT-CLJ0236-4938 & 0.334 & 4.0 & 0.19$^{+0.06}_{-0.07}$ & 0.035$^{+0.006}_{-0.004}$ \\
SPT-CLJ0243-5930 & 0.635 & 4.8 & 0.17$^{+0.04}_{-0.04}$ & 0.016$^{+0.003}_{-0.003}$ \\
SPT-CLJ0252-4824 & 0.330 & 4.1 & 0.97$^{+0.34}_{-0.28}$ & 0.034$^{+0.008}_{-0.014}$ \\
SPT-CLJ0256-5617 & 0.580 & 4.3 & 1.53$^{+0.72}_{-0.35}$ & 0.068$^{+0.013}_{-0.018}$ \\
SPT-CLJ0304-4401 & 0.460 & 10.4 & 0.50$^{+0.07}_{-0.08}$ & 0.044$^{+0.006}_{-0.003}$ \\
SPT-CLJ0304-4921 & 0.392 & 6.2 & 0.11$^{+0.02}_{-0.03}$ & 0.010$^{+0.002}_{-0.002}$ \\
SPT-CLJ0307-5042 & 0.550 & 4.4 & 0.08$^{+0.03}_{-0.04}$ & 0.015$^{+0.003}_{-0.003}$ \\
SPT-CLJ0307-6225 & 0.581 & 5.7 & 2.46$^{+0.22}_{-0.28}$ & 0.136$^{+0.034}_{-0.011}$ \\
SPT-CLJ0310-4647 & 0.850 & 10.6 & 0.07$^{+0.06}_{-0.08}$ & 0.014$^{+0.003}_{-0.004}$ \\
SPT-CLJ0324-6236 & 0.730 & 4.1 & 0.02$^{+0.03}_{-0.03}$ & 0.013$^{+0.003}_{-0.003}$ \\
SPT-CLJ0330-5228 & 0.442 & 4.4 & 0.08$^{+0.03}_{-0.07}$ & 0.040$^{+0.005}_{-0.007}$ \\
SPT-CLJ0334-4659 & 0.450 & 5.5 & 0.16$^{+0.04}_{-0.03}$ & 0.018$^{+0.003}_{-0.002}$ \\
SPT-CLJ0346-5439 & 0.530 & 4.1 & 0.12$^{+0.02}_{-0.04}$ & 0.023$^{+0.004}_{-0.004}$ \\
SPT-CLJ0348-4515 & 0.358 & 3.9 & 0.07$^{+0.08}_{-0.13}$ & 0.016$^{+0.004}_{-0.005}$ \\
SPT-CLJ0352-5647 & 0.670 & 3.7 & 0.08$^{+0.05}_{-0.07}$ & 0.017$^{+0.003}_{-0.005}$ \\
SPT-CLJ0406-4805 & 0.590 & 4.0 & 0.37$^{+0.14}_{-0.19}$ & 0.054$^{+0.010}_{-0.008}$ \\
SPT-CLJ0411-4819 & 0.422 & 6.0 & 0.82$^{+0.04}_{-0.04}$ & 0.057$^{+0.001}_{-0.002}$ \\
SPT-CLJ0417-4748 & 0.590 & 4.9 & 0.05$^{+0.02}_{-0.03}$ & 0.010$^{+0.002}_{-0.002}$ \\
SPT-CLJ0426-5455 & 0.630 & 3.8 & 0.22$^{+0.27}_{-0.34}$ & 0.046$^{+0.010}_{-0.010}$ \\
SPT-CLJ0438-5419 & 0.421 & 10.3 & 0.25$^{+0.04}_{-0.03}$ & 0.018$^{+0.001}_{-0.001}$ \\
SPT-CLJ0441-4855 & 0.790 & 4.0 & 0.06$^{+0.03}_{-0.02}$ & 0.013$^{+0.003}_{-0.002}$ \\
SPT-CLJ0446-5849 & 1.182 & 3.6 & 0.21$^{+0.36}_{-0.90}$ & 0.059$^{+0.011}_{-0.012}$ \\
SPT-CLJ0449-4901 & 0.792 & 5.1 & 0.27$^{+0.07}_{-0.10}$ & 0.046$^{+0.007}_{-0.006}$ \\
SPT-CLJ0456-5116 & 0.570 & 4.4 & 0.11$^{+0.05}_{-0.08}$ & 0.023$^{+0.005}_{-0.004}$ \\
SPT-CLJ0509-5342 & 0.461 & 5.7 & 0.07$^{+0.02}_{-0.02}$ & 0.008$^{+0.002}_{-0.002}$ \\
SPT-CLJ0516-5430 & 0.295 & 12.4 & 0.18$^{+0.07}_{-0.16}$ & 0.055$^{+0.003}_{-0.004}$ \\
SPT-CLJ0522-4818 & 0.296 & 3.7 & 0.22$^{+0.08}_{-0.11}$ & 0.016$^{+0.003}_{-0.004}$ \\
SPT-CLJ0528-5300 & 0.768 & 2.4 & 0.07$^{+0.04}_{-0.08}$ & 0.024$^{+0.005}_{-0.007}$ \\
SPT-CLJ0533-5005 & 0.881 & 1.6 & 0.21$^{+0.16}_{-0.24}$ & 0.049$^{+0.011}_{-0.009}$ \\
SPT-CLJ0542-4100 & 0.642 & 5.7 & 0.55$^{+0.21}_{-0.20}$ & 0.041$^{+0.004}_{-0.005}$ \\
SPT-CLJ0546-5345 & 1.066 & 5.0 & 0.10$^{+0.03}_{-0.04}$ & 0.022$^{+0.003}_{-0.004}$ \\
SPT-CLJ0551-5709 & 0.423 & 3.0 & 0.88$^{+0.12}_{-0.16}$ & 0.086$^{+0.007}_{-0.006}$ \\
SPT-CLJ0555-6406 & 0.270 & 5.6 & 0.36$^{+0.09}_{-0.13}$ & 0.039$^{+0.004}_{-0.003}$ \\
SPT-CLJ0559-5249 & 0.609 & 5.5 & 0.39$^{+0.10}_{-0.08}$ & 0.041$^{+0.006}_{-0.007}$ \\
SPT-CLJ0616-5227 & 0.684 & 4.5 & 0.34$^{+0.10}_{-0.06}$ & 0.031$^{+0.005}_{-0.004}$ \\
SPT-CLJ0655-5234 & 0.500 & 4.8 & 0.30$^{+0.12}_{-0.21}$ & 0.018$^{+0.005}_{-0.007}$ \\
SPT-CLJ0658-5556 & 0.296 & 18.2 & 2.41$^{+0.01}_{-0.01}$ & 0.020$^{+0.000}_{-0.000}$ \\
SPT-CLJ2011-5725 & 0.279 & 2.6 & 0.11$^{+0.02}_{-0.02}$ & 0.006$^{+0.001}_{-0.001}$ \\
SPT-CLJ2031-4037 & 0.330 & 7.4 & 0.25$^{+0.04}_{-0.04}$ & 0.017$^{+0.001}_{-0.002}$ \\
SPT-CLJ2034-5936 & 0.946 & 4.4 & 0.05$^{+0.05}_{-0.23}$ & 0.014$^{+0.003}_{-0.005}$ \\
SPT-CLJ2035-5251 & 0.520 & 4.1 & 0.07$^{+0.13}_{-0.35}$ & 0.039$^{+0.011}_{-0.010}$ \\
SPT-CLJ2043-5035 & 0.723 & 5.1 & 0.11$^{+0.02}_{-0.02}$ & 0.009$^{+0.002}_{-0.002}$ \\
SPT-CLJ2106-5844 & 1.132 & 8.1 & 0.20$^{+0.04}_{-0.03}$ & 0.022$^{+0.002}_{-0.002}$ \\
SPT-CLJ2135-5726 & 0.427 & 4.5 & 0.10$^{+0.04}_{-0.08}$ & 0.013$^{+0.003}_{-0.004}$ \\
SPT-CLJ2145-5644 & 0.480 & 5.7 & 0.06$^{+0.03}_{-0.04}$ & 0.011$^{+0.003}_{-0.003}$ \\
SPT-CLJ2146-4633 & 0.933 & 3.5 & 0.05$^{+0.05}_{-0.18}$ & 0.026$^{+0.007}_{-0.009}$ \\
SPT-CLJ2148-6116 & 0.571 & 5.0 & 0.06$^{+0.03}_{-0.07}$ & 0.018$^{+0.004}_{-0.004}$ \\
SPT-CLJ2218-4519 & 0.650 & 4.9 & 0.08$^{+0.05}_{-0.08}$ & 0.027$^{+0.005}_{-0.006}$ \\
SPT-CLJ2222-4834 & 0.652 & 4.0 & 0.09$^{+0.03}_{-0.05}$ & 0.015$^{+0.003}_{-0.004}$ \\
SPT-CLJ2232-5959 & 0.594 & 3.3 & 0.10$^{+0.03}_{-0.03}$ & 0.008$^{+0.002}_{-0.002}$ \\
SPT-CLJ2233-5339 & 0.480 & 6.4 & 0.29$^{+0.12}_{-0.11}$ & 0.026$^{+0.005}_{-0.003}$ \\
SPT-CLJ2236-4555 & 1.161 & 2.7 & 0.90$^{+0.18}_{-0.19}$ & 0.063$^{+0.007}_{-0.009}$ \\
SPT-CLJ2245-6206 & 0.580 & 5.1 & 0.87$^{+0.19}_{-0.19}$ & 0.113$^{+0.007}_{-0.010}$ \\
SPT-CLJ2248-4431 & 0.351 & 16.7 & 0.21$^{+0.03}_{-0.02}$ & 0.006$^{+0.001}_{-0.000}$ \\
SPT-CLJ2258-4044 & 0.864 & 4.0 & 0.31$^{+0.15}_{-0.24}$ & 0.036$^{+0.007}_{-0.004}$ \\
SPT-CLJ2259-6057 & 0.750 & 4.3 & 0.11$^{+0.03}_{-0.03}$ & 0.020$^{+0.002}_{-0.003}$ \\
SPT-CLJ2301-4023 & 0.730 & 2.4 & 0.61$^{+0.14}_{-0.12}$ & 0.028$^{+0.003}_{-0.003}$ \\
SPT-CLJ2306-6505 & 0.530 & 5.6 & 0.05$^{+0.04}_{-0.12}$ & 0.029$^{+0.006}_{-0.008}$ \\
SPT-CLJ2325-4111 & 0.358 & 8.2 & 0.36$^{+0.09}_{-0.10}$ & 0.040$^{+0.004}_{-0.005}$ \\
SPT-CLJ2331-5051 & 0.576 & 4.6 & 0.14$^{+0.03}_{-0.05}$ & 0.034$^{+0.002}_{-0.003}$ \\
SPT-CLJ2332-5053 & 0.560 & 5.3 & 4.10$^{+0.48}_{-0.43}$ & 0.090$^{+0.005}_{-0.007}$ \\
SPT-CLJ2335-4544 & 0.570 & 5.4 & 0.04$^{+0.03}_{-0.07}$ & 0.015$^{+0.004}_{-0.004}$ \\
SPT-CLJ2337-5942 & 0.775 & 5.9 & 0.09$^{+0.03}_{-0.05}$ & 0.016$^{+0.002}_{-0.003}$ \\
SPT-CLJ2341-5119 & 1.003 & 5.8 & 0.07$^{+0.02}_{-0.03}$ & 0.017$^{+0.002}_{-0.003}$ \\
SPT-CLJ2342-5411 & 1.075 & 1.9 & 0.08$^{+0.03}_{-0.04}$ & 0.011$^{+0.002}_{-0.004}$ \\
SPT-CLJ2344-4243 & 0.596 & 11.9 & 0.03$^{+0.00}_{-0.00}$ & 0.002$^{+0.000}_{-0.000}$ \\
SPT-CLJ2345-6405 & 0.962 & 4.9 & 0.33$^{+0.19}_{-0.53}$ & 0.045$^{+0.008}_{-0.007}$ \\
SPT-CLJ2352-4657 & 0.783 & 4.0 & 0.02$^{+0.03}_{-0.04}$ & 0.020$^{+0.004}_{-0.005}$ \\
SPT-CLJ2355-5055 & 0.320 & 3.9 & 0.07$^{+0.02}_{-0.02}$ & 0.013$^{+0.003}_{-0.003}$ \\
SPT-CLJ2359-5009 & 0.775 & 2.9 & 0.24$^{+0.08}_{-0.14}$ & 0.016$^{+0.005}_{-0.006}$ \\
\hline
\end{longtable}       
\end{center}

We consider the following pairs of subsamples in our comparisons:

\begin{description}
\item [A)] \textbf{400d low-z vs SPT low-z} ($z < 0.6$; 27 \& 50 clusters 
respectively).  These are maximally similar catalogs in 
mass-redshift space, allowing for the cleanest test of morphology differences between X-ray- and SZ-selected cluster samples. Within this redshift range, we find no correlation between mass and \Aphot\ (Pearson r $=$ 0.11) for simulated clusters, suggesting that poor overlap in mass between the two samples will not drive any result.

\item [B)] \textbf{400d vs SPT, all redshifts} (36 \& 90 clusters). Allows a comparison of cluster morphologies in the complete SZ and 
    X-ray selected catalogs, ignoring that these samples have different redshift and mass ranges.

\item [C)] \textbf{SPT low-z vs SPT high-z} ($z < 0.6$, $z>0.6$; 50 \& 40 clusters, respectively). This is the cleanest 
test for substructure evolution, since the high-$z$ SPT-selected clusters can be considered the progenitors of the low-$z$ SPT-selected systems \citep[see Figure 7 of][]{mcdonald14c}. Given the relatively small change in angular size between the low-$z$ and high-$z$ systems, we expect there to be minimal redshift-dependent selection biases between these two subsamples.

\item [D)] \textbf{400d low-z vs SPT high-z} ($z < 0.6$, $z>0.6$; 27 \& 40 clusters, respectively). This is a complementary test for 
substructure evolution, which has the underlying evolution convolved with redshift-dependent, mass-dependent, and method-dependent selection bias. 
Naively, one might assume that both SZ-selected clusters and high-redshift clusters are 
more likely to be mergers, because mergers can cause temporarily increased pressure, and the merger rate is higher at early times \citep[e.g.,][]{fakhouri10}. This would lead to the high-$z$, SPT-selected clusters being significantly more disturbed than the low-$z$, X-ray-selected clusters, if these statements are true. A lack of difference in morphology between high-z 
SZ-selected and low-z X-ray selected clusters would indicate that the combination of these effects is insignificant.

\item [E)] \textbf{400d+SPT low-z vs 400d+SPT high-z} ($z<0.6$, $z>0.6$; 75 \& 49 clusters, respectively). 
If selection criteria are indeed not important, this test increases the statistical 
power of the substructure evolution test due to increased number of clusters in 
the combined samples.
    
\item [F,G,H)] \textbf{SPT+400d vs simulations}. For these comparisons, we select real and simulated clusters within the redshift intervals of $z=0.33\pm0.1$, $z=0.66 \pm 0.1$, and $z=0.99 \pm 0.15$.

\end{description}

\section{Results}
\label{sec:results}
\subsection{X-ray --- SZ Comparisons}

\bfc
    \plotone{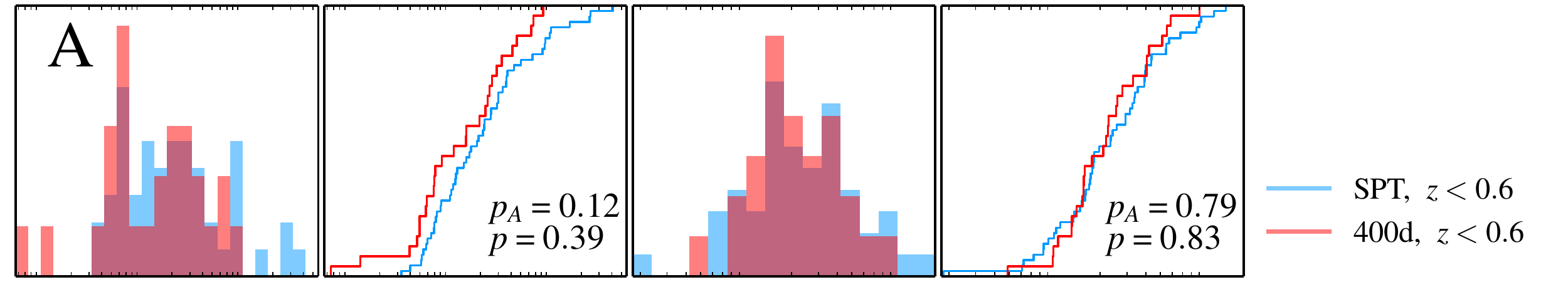}
    \plotone{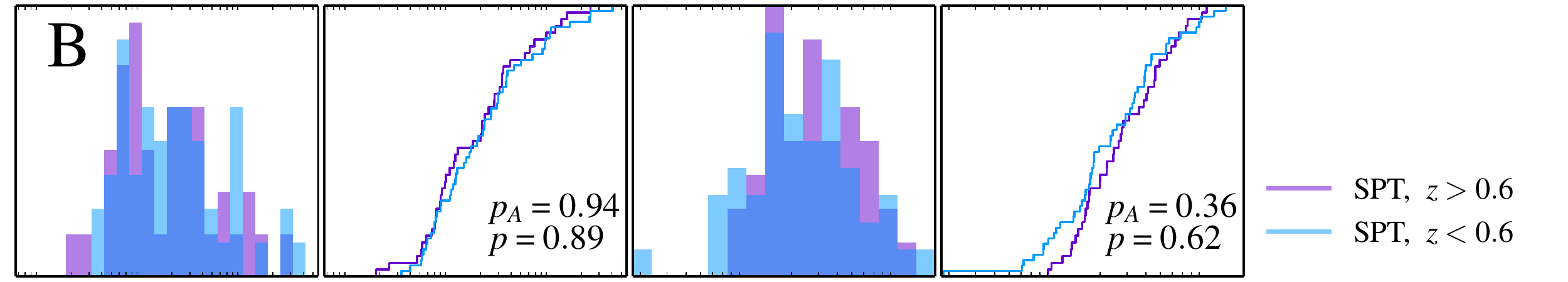}
    \plotone{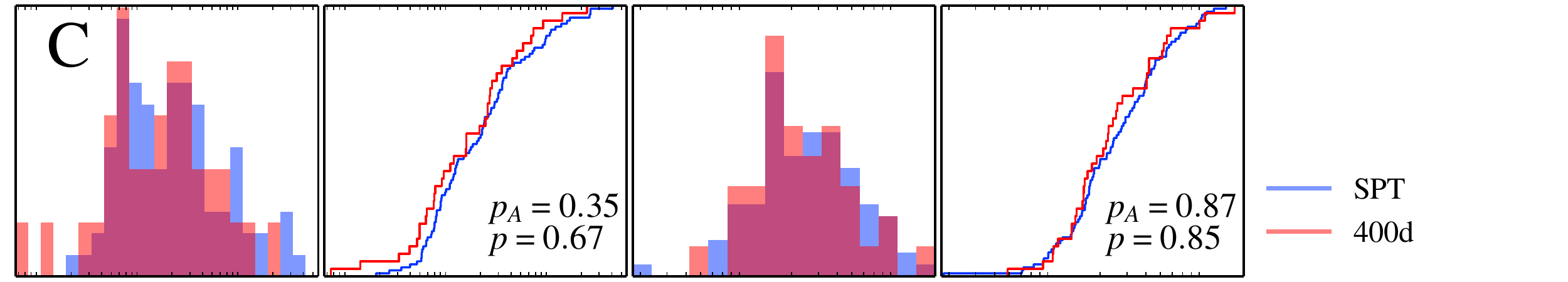}
    \plotone{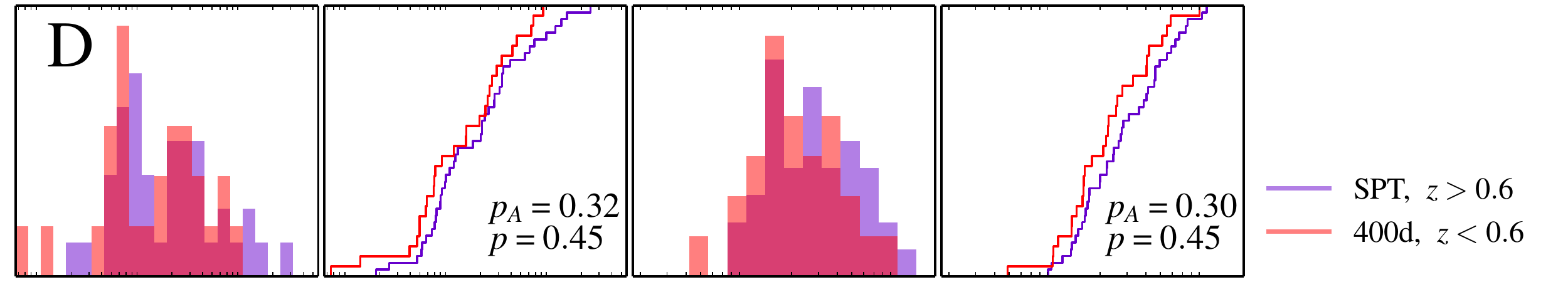}
    \plotone{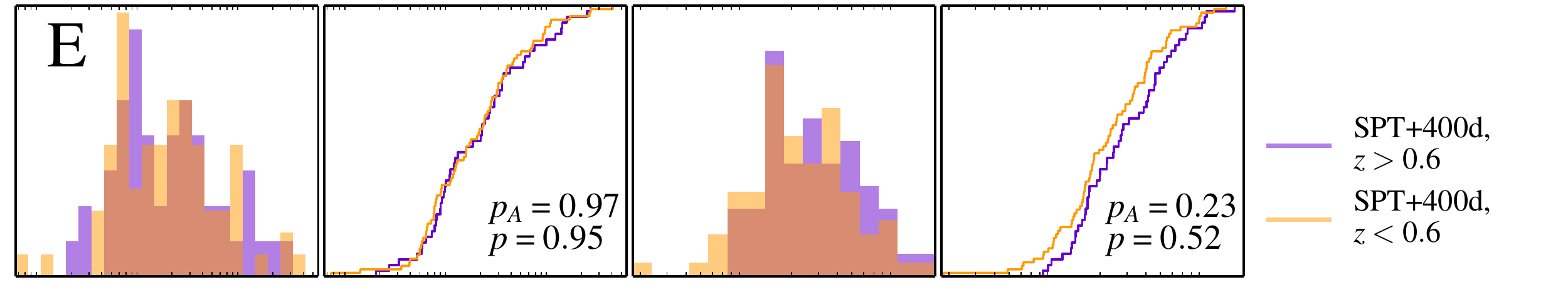}

    \plotone{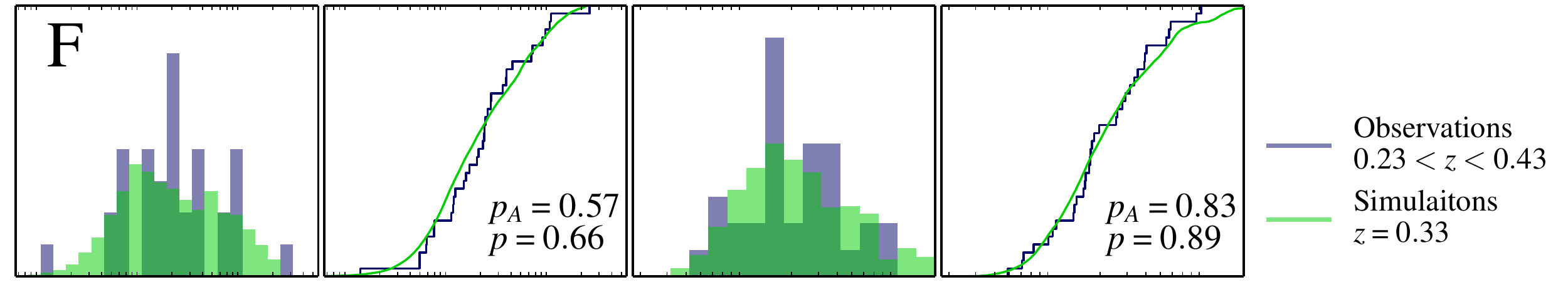}
    \plotone{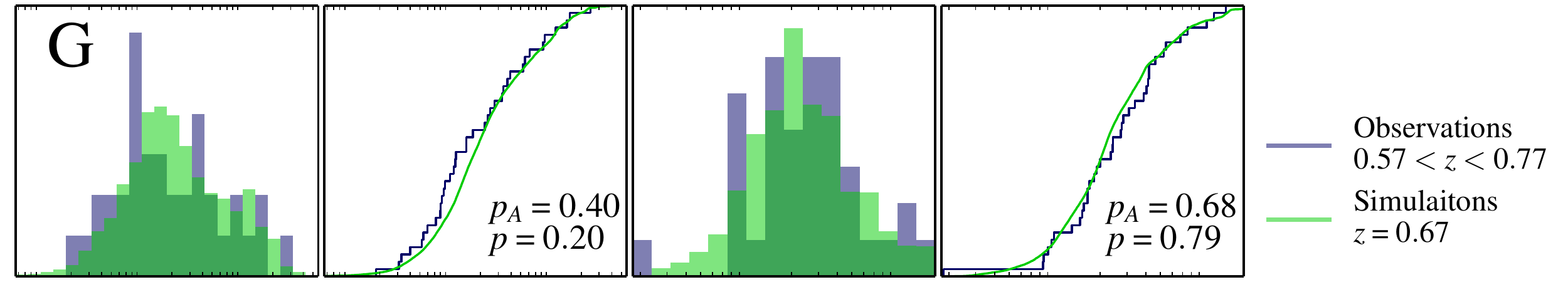}
    \plotone{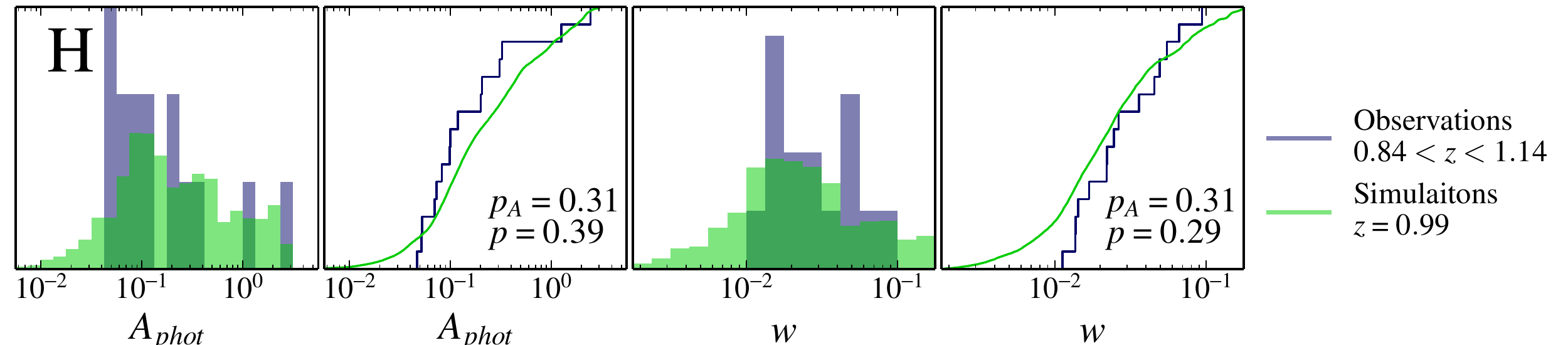}
    \caption{Distribution of morphological statistics for various partitions of 
    clusters in the 400d and SPT catalogs. In each row, the left two plots show the histograms 
and cumulative distribution functions for \Aphot, and the right two plots show the same for 
$w$. The $p$-value, based on the KS and AD tests, of the hypothesis that the substructure statistic values come 
from the same distribution are shown in each plot. High values of $p$ indicate 
that the distributions are similar, low values of $p$ indicate that the 
distributions are different. It is a common practice to exclude the null 
hypothesis at $p < 0.05/ (\textrm{number of tests being conducted})$.
The rightmost column explains what samples are being compared.  See 
Sec.~\ref{sec:methods} for the motivation of the subsamples to be compared and 
Sec.~\ref{sec:results} for the description of the results.  }

\label{fig:comparisons}
\efc

Fig.~\ref{fig:comparisons}A shows histograms and cumulative distributions of 
\Aphot\ and \w\ for 400d and SPT subsamples with $z<0.6$. In general, the distribution of substructure statistics is closer to log-normal than normal.
High $p$-values ($p>0.1$) of 
the KS and AD statistics for \Aphot\ and \w\ indicate that the two samples (400d and SPT; $z<0.6$) are indistinguishable in terms of their X-ray morphology. Under the 
assumption that mergers are characterized by increased values of \Aphot\ and \w, 
this means that the fraction of merging systems detected by SPT is 
similar to the amount of merging systems detected by their X-ray emission.  
Assuming that ongoing mergers can be identified as having \Aphot\ $>$ 0.6, following Figure \ref{fig:merger} \citep[see also Fig.\ 7 from][]{Nurgaliev13}, we find a low-$z$ merging fraction of $20^{+7}_{-4}$\% and $11^{+9}_{-4}$\% in the SPT and 400d samples, respectively, where the uncertainty range is the 1$\sigma$ binomial population confidence interval \citep{Cameron11}.
 For comparison, \cite{mann12} find a merger fraction of $24^{+5}_{-4}$\% for a sample of 79 X-ray selected clusters spanning the same redshift range ($0.25 < z < 0.6$).

\begin{figure*}[ht]
    \epsscale{1.2}
    \plotone{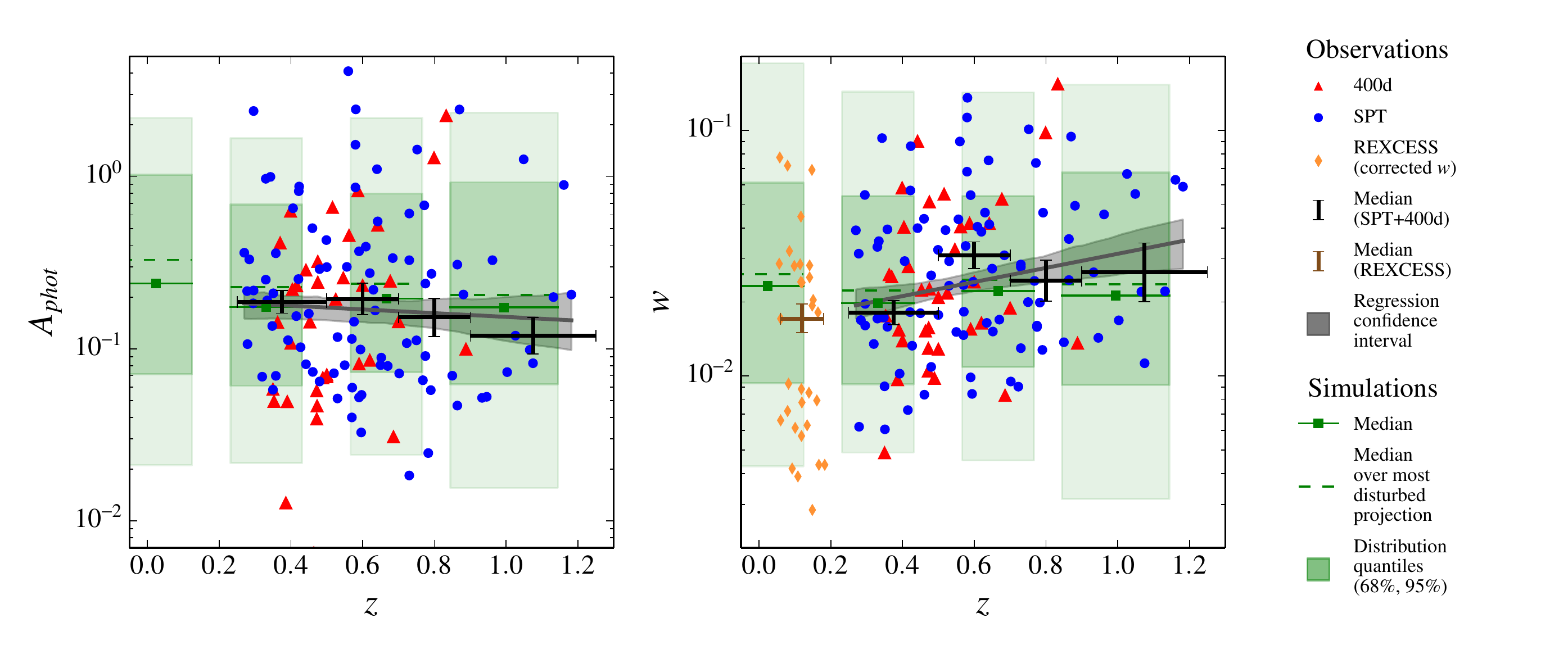}
    \caption{Redshift evolution of \Aphot\ and \w. Red triangles and blue circles represent 400d and SPT clusters 
respectively. Black error bars show the median values in four redshift bins ([0.25--0.5], [0.5--0.7], [0.7--0.9], [0.9--1.2]). We measure Pearson R coefficients for the left and right panels of 0.02 and 0.19, respectively, indicating a lack of a statistically significant correlation.
The error bars on the median value are obtained by the Median Absolute  
Devitation method and demonstrate that there is no significant difference in the median 
value between bins. The slopes, derived by a simple linear regressions of \Aphot\ and \w\ with $z$, are 
consistent with no redshift evolution at $\sim$0.5$\sigma$ and $\sim$2$\sigma$ 
respectively. The shaded green regions show the 68\% and 95\% ranges for \Aphot\ and \w\ in simulated clusters, with the medians shown as green squares. The orange circles in the 
right panel show the values of $w$ for the REXCESS sample presented in 
\citet{Bohringer10}, accordingly corrected for the different \w\ definition used in this study.
}
    \label{fig:z_evol}
\end{figure*}

We next compare clusters at low and high redshift, subject 
to the same selection criteria. We consider $z < 0.6$  and $z > 0.6$ subsamples 
of the parent SPT sample only, because the 400d sample only has 9 clusters at $z>0.6$.
Fig.~\ref{fig:comparisons}B shows that the distributions of both \Aphot\ and \w\ 
for low-z and high-z systems are indistinguishable ($p>0.3$). Using the same criterion as above, we find that $20^{+7}_{-4}$\% and $18^{+8}_{-4}$\% of clusters at low-$z$ and high-$z$, respectively, are identified as ongoing mergers.

Given that the SZ-selected clusters demonstrate weak, if any, redshift evolution in their morphology, we next consider the full 400d and SPT catalogs, without any redshift restrictions. 
Fig.~\ref{fig:comparisons}C shows 
distribution functions for these two samples, which are still indistinguishable ($p>0.3$) for both \Aphot\ and \w. The fraction of systems with \Aphot\ $>$ 0.6 over the full redshift range is $19^{+5}_{-3}$\% and $14^{+8}_{-4}$\% for the SPT and 400d samples, respectively.

If there was only a weak dependence of X-ray morphology on both selection and redshift, we might not have sufficient statistical power to detect such dependencies with the tests described above. In an effort to maximize the effects of these two potential biases, we compare the low-$z$ X-ray sample to the high-$z$ SZ sample. Fig.~\ref{fig:comparisons}D shows the result of this comparison, demonstrating that there is no statistically significant difference in the distribution of X-ray morphologies between these two extreme subsamples ($p>0.3$). This means that, if clusters are, on average, more morphologically disturbed in SZ-selected clusters \emph{and} at high redshift, the combined effects on the measured disturbed fraction are small ($\lesssim$10\%).

As a final test, we make the assumption that there is no morphological bias due to selection, and combine the 400d and SPT samples to maximize our ability to detect redshift dependence. Fig.~\ref{fig:comparisons}E demonstrates that there is no measurable redshift dependence even when the two samples are combined, for both \Aphot\ and \w\ ($p>0.2$). In this combined sample, we find that the fraction of clusters with \Aphot\ $>$ 0.6 is $17^{+5}_{-3}$\% and $19^{+7}_{-4}$\% for the low-$z$ and high-$z$ subsets, respectively.

In summary, we find no statistically significant dependence on either the distribution of X-ray morphologies or the fraction of clusters classified as morphologically-disturbed with redshift or selection method.

\subsection{Data --- Simulations Comparisons}

Figures~\ref{fig:comparisons}F, G, H show comparisons between observed and 
simulated clusters at three redshifts: $z=0.33$, 0.66 and 0.99. We note a 
remarkable agreement in the X-ray morphology between simulations and 
observations. The lowest measured p-value is 0.20 (for KS test on \Aphot\ at $z=0.67$). Given that we have made 16 individual comparisons, we require $p < 0.003$ to reject the null hypothesis that these two distributions come from the same parent distribution. From the similarity in the data and simulations, we can arrive at two conclusions. First, the lack of evolution in X-ray morphology is observed in both simulations and observations, suggesting that this is not due to a selection bias. Secondly, the observed morphology is relatively insensitive to complex physics (e.g., cooling, AGN feedback, etc.), and appears to be primarily driven by gravitational processes (i.e., mergers), which simulations adequately describe.

In Figure~\ref{fig:z_evol}, we show the X-ray morphology, as quantified by \Aphot\ and \w, as a function of 
redshift for the 400d and SPT samples, as well as for the \textit{Omega500} simulations.  For \w, we compare to the low-$z$ REXCESS cluster sample \citep{Bohringer10}. Our definition of \w\ \citep{Nurgaliev13} is slightly different than that 
used by \citet{Bohringer10} --- to make our results comparable we compute \w\ for all clusters in the 400d and SPT samples using both methods and found that scaling by a factor of 1.5 brings the two into excellent agreement. After the applied correction, the range and median value of \w\ in the 
study of \citet{Bohringer10} are similar to those for low- and high-z clusters in the 400d and SPT samples. 

\begin{figure*}[htb]
    \epsscale{1.2}
    \plotone{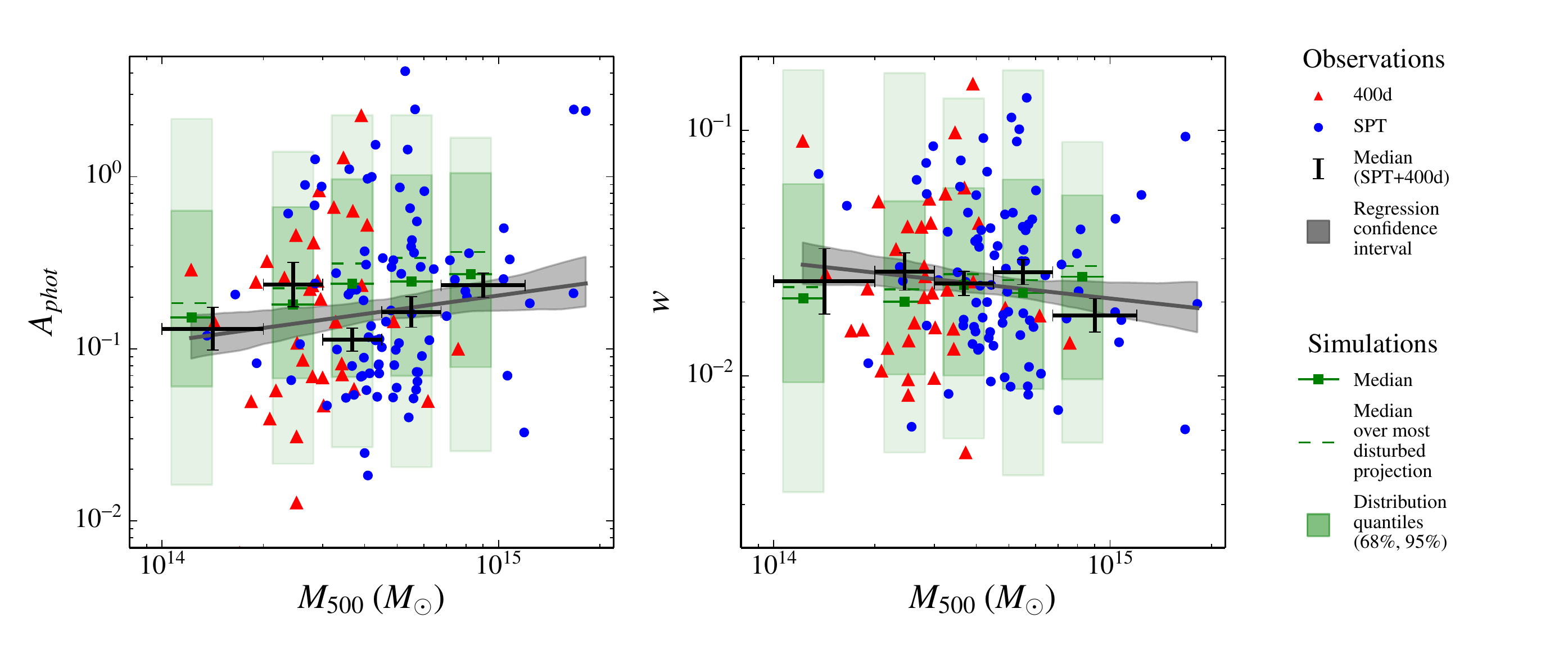}
    \caption{Similar to Figure \ref{fig:z_evol}, but now considering cluster mass rather than redshift. We measure Pearson R coefficients for the left and right panels of 0.19 and -0.11, respectively, indicating a lack of a statistically significant correlation. There is no significant correlation between cluster asymmetry and mass, regardless of the asymmetry estimator used or whether we consider real or simulated clusters.}
    \label{fig:zm_morph}
\end{figure*}

Figure~\ref{fig:z_evol} demonstrates that the amount of substructure, as quantified by 
\w\ and \Aphot, is remarkably similar for four different samples: 400d, SPT, 
REXCESS and the \textit{Omega500} simulations. Further, there is no statistically significant redshift evolution over the combined samples.
We note that the mild (2$\sigma$) redshift evolution measured in \w\ is consistent with our earlier findings that \w\ is biased high for low-quality X-ray data \citep{Nurgaliev13}. Given that the most distant clusters have higher background and fewer net counts than their low-$z$ counterparts, we expect them to be biased high in \w.

In Figure \ref{fig:zm_morph}, we show the same measurements of \Aphot\ and \w\ as in Figure \ref{fig:z_evol}, but now as a function of cluster mass. We find no statistically significant dependence of cluster morphology on cluster mass, over the mass range of $10^{14}$--$10^{15}$ M$_{\odot}$, for both simulated and real clusters. Importantly, the fact that there is no mass trend in the simulated clusters means that we are justified in comparing the simulated and real clusters despite the fact that their mass distributions are not identical.

\section{Discussion}
\label{sec:discussion}

\subsection{X-ray, SZ selection biases}

We have demonstrated in Figs.\ \ref{fig:comparisons}, \ref{fig:z_evol}, and \ref{fig:zm_morph} that there is no measurable difference between the distribution of X-ray morphology in the 400d X-ray-selected sample and the SPT SZ-selected sample. Further, both of these selections appear to be unbiased with respect to simulations, suggesting that we are probing the full population of massive galaxy clusters. These results are consistent with work by \cite{mantz15} who found no difference in the relaxed fraction between the 400d and SPT samples, and with \cite{sifon13,sifon16} who used dynamical tracers of substructure to show that the dynamical state of SZ-selected clusters from ACT were consistent with those of simulated massive clusters.

There persists a misconception that SZ-selected clusters are, on average, more often mergers than X-ray-selected clusters. We propose that this perceived bias stems from two facts pertaining to \emph{Planck}-selected clusters that were summarized briefly in the introduction. First, \emph{Planck} has a significantly more extended PSF ($\sim$7$^{\prime}$) compared to ground-based arcminute-resolution SZ experiments such as SPT and the Atacama Cosmology Telescope \citep[ACT;][]{swetz11}. This means that, at high-$z$, \emph{Planck} does not resolve close pairs (or triplets) of galaxy clusters (e.g.  PLCKG214.6+37.0), thereby capturing an ``inflated'' SZ signal, 
where an instrument with a smaller beam such as SPT or ACT would see multiple independent 
systems with lower individual significance. Because the SPT beam ($\sim$1$^{\prime}$) is matched to the angular 
size of rich clusters at $z>0.3$, it is unlikely that the SPT sample contains similar 
blended systems. Secondly, several major mergers in the \emph{Planck} catalog received a great deal of initial attention, due to the fact that there were very few such systems previously known (i.e., triple clusters). However, the detection of a few dozen previously unknown mergers compared to more 
than 800 confirmed objects in \emph{Planck's} catalog is not a statistically 
significant indication of morphological bias in SZ-selected cluster samples.

The results presented here are also consistent with recent work by \cite{Lin15}, who showed that the bias due to the presence or lack of a cool core is small ($<$1\%) for the majority of systems, with exceptionally rare systems like the Phoenix cluster \citep{mcdonald12c} having biases as high as 10\%. Applying the results of \cite{Lin15} to the HIFLUGCS sample of low-$z$ clusters \citep{reiprich02,Vikhlinin07}, we estimate that 1 in $\sim$100 clusters has an SZ bias as high as $\sim$10\%. This bias is strongly redshift dependent, due to the combined effects of cool cores filling a smaller fraction of the beam at high-$z$, and being in general less cuspy at early times \citep{mcdonald13b, mantz15}. As such, we expect the fraction of cool cores in the SPT-selected cluster sample to be nearly representative of the true underlying population. Convolving this very weak bias with the noisy correlation between X-ray morphology and the presence or lack of a cool core, we do not expect a cool core bias to drive a statistically significant difference in morphology between our X-ray- and SZ-selected samples.

We conclude that the common misconception that SZ-selected samples contain a non-representative fraction of mergers stems from some combination of these two points. We find no morphological bias in an SPT-selected sample, and would not expect any similarly-selected samples to be biased either \cite[see also][]{motl05,Lin15}.

\subsection{Evolution of substructure with redshift}

In \S4.1 we demonstrated that low- and high-redshift systems show the same amount of 
substructure. This is somewhat surprising --- in the standard growth of structure 
scenario, the fraction of disturbed clusters increases with increasing redshift.  
Additionally, these findings are in contradiction with 
some earlier works, specifically \citet{jeltema05} and \citet{andersson09}, which both rely on power ratios to quantify morphology.  
Both \citet{Nurgaliev13} and \citet{weissmann13b} have shown that shot noise 
can strongly influence the measured power ratios. This may explain the results found by \citet{andersson09}, who did not apply any shot noise correction.


One possible explanation for the lack of observed X-ray morphology evolution may be 
that there is not a one-to-one correspondence between substructure statistics 
such as \w\ or \PR and the dynamical state of the cluster. Indeed, simulations have shown that 
substructures statistics may vary significantly on short time scales during 
cluster mergers \citep{Hallman11, Poole06, OHara06}. For example, simulations by 
\citet{OHara06} and \citet{Poole06} show that \w\ can easily vary in the range 
$0.01<w<0.1$ over a fraction of characteristic merger time. In addition \w\ shows similarly big fluctuations as a 
function of line of sight.

Another possibility (as pointed out by \citealt{weissmann13b}) is that earlier 
studies that used power ratios might not account correctly for the insufficient 
photon statistics of high-z clusters. For example, \citet{jeltema05} analyzed 
redshift evolution using a relatively small sample of 40 clusters divided into 
low and high redshift subsamples. The low redshift  contained 26 clusters with 
$z<0.5$ and $\langle z \rangle = 0.24$. The high redshift sample contained 14 
clusters with $z>0.5$ and $\langle z \rangle = 0.71$.  They used the power ratio 
method and found the amount of substructure significantly different between the 
subsamples as measured by $P_3/P_0$.  They also fit the $P_3/P_0 - z$ 
relation and found the slope to be positive with high ($p \approx 0.005$) 
significance. Their results are surprising in light of new studies of the 
properties of the power-ratios method. Both \citet{Nurgaliev13} and 
\citet{weissmann13b} find that $P_3/P_0$ are fully consistent with zero for a 
majority of high-redshift clusters due to insufficient quality of observations.

\citet{weissmann13b} performed a similar study using the same high-z 400d 
sample and a subsample of the SPT sample used in \citet{Andersson11}, combined with a low-z sample published in \citet{weissmann13b}. 
They quantified morphology via centroid shifts and power ratios, finding a result consistent with no 
redshift evolution. 
%
\citet{weissmann13b} emphasized bringing both high-redshift and 
low-redshift subsamples to the same quality of observations. To achieve that, 
they artificially degrade higher-quality observations of the low-redshift 
sample.  This is not necessary in our analysis because 1) the observations of 
both samples were targeted for 2000 counts per cluster and 2) as shown in 
\citet{Nurgaliev13} \w\ is not sensitive to number of counts above $\sim$ 1000 
counts (this is also confirmed in \citealt{weissmann13b}) and \Aphot\ has even 
better stability than \w\ with respect to the number of X-ray counts. Indeed, the \Aphot\ quantity was derived explicitly to avoid any bias due to data quality.
The work presented here extends on that of \citet{weissmann13b} by including a larger number of distant, SPT-selected
clusters, by splitting high-redshift clusters into multiple selection (X-ray, SZ, simulation) bins, and by using a 
new substructure statistic, \Aphot.

A similar study was also conducted by \cite{mantz15}, utilizing data from both the SPT and 400 deg$^2$ surveys along with data for RASS and \emph{Planck} clusters. A direct comparison to the results of this study is not straightforward, because it was focused on the ``relaxed fraction'' using a conservative estimator for relaxedness, while this work focuses on the evolution of the full distribution of morphologies. Nonetheless, \cite{mantz15} find that the relaxed fraction does not evolve significantly between $z\sim0$ and $z\sim1$, and that there is no statistically significant difference between the morphologies of SPT-selected and 400d-selected clusters, consistent with this work.

Although we do not find evidence for a change in the amount of substructure with redshift based on a robust non-parametric statistical test, one can speculate about the fact that there are a larger fraction of systems in the low-$z$ subsample with unusually low \w\ and \Aphot, compared to the high-$z$ subsample.  
Based on the KS and AD tests, and a robust comparison of the medians in different subsamples, we cannot call this a significant effect, however. Both subsamples could be drawn from the same underlying distribution.

\subsection{Comparison with simulated clusters}
\label{sec:simulations}

In general, the simulated clusters studied here \citep{Nagai07} look quantitatively similar to the observed clusters. Distributions of both \w\ and \Aphot\ for real and simulated clusters of the same mass and redshift were statistically indistinguishable. Given that these simulations did not include complex astrophysics, such as AGN feedback, we can conclude that, while the development of dense cool cores is sensitive to the specifics of the feed-back prescriptions \citep[e.g.,][]{Gaspari14}, overall asymmetry in the ICM is not.
This is perhaps not surprising, given that gravitational processes (i.e., mergers) are the dominant source of asymmetry in the ICM.

We can conclude from this work that no more complicated physics is necessary to broadly match the morphology of real and simulated clusters than was included in the $Omega500$ simulations. The combined effects of physical processes including AGN feedback, non-ideal inviscid fluids, and cosmic rays are minimal, and do not significantly bias the observed morphology.

\section{Conclusions}
\label{sec:conclusions}

Using samples of 36 X-ray selected clusters from the 400 deg$^2$ ROSAT survey, 91 SZ-selected clusters from the South Pole Telescope 2500 deg$^2$ survey, and 85 simulated clusters from the \emph{Omega}500 simulations, 
all observed (or mock observed) to roughly equal depth with the \emph{Chandra X-ray Observatory}, we investigated whether these samples have any bias towards 
cluster morphological type, and whether high-redshift clusters are more 
disturbed than their low-redshift counterparts. We considered two well-defined
substructure statistics and tested for statistically significant differences in their 
distributions between different subsamples. In the mass and redshift range studied, we find no evidence for a statistically significant difference in the X-ray morphologies of clusters selected via X-ray or SZ, or at low or high redshift. Further, we found that simulated clusters had quantitatively similar morphology to X-ray- and SZ-selected systems, considering only the asymmetry of the hot gas (i.e., ignoring central cusps).

Our results demonstrate that there is no significant bias for or against preferentially selecting mergers in high resolution ($\sim$1$^{\prime}$) SZ surveys. For SZ surveys with larger beam size (e.g., \emph{Planck}), morphological biases may exist due to the fact that multiple clusters or extended structures can contribute to the integrated signal.

\section*{Acknowledgements} 

Much of this work was enabled by generous GTO contributions from Steve Murray, and was in progress at the time of his untimely death in 2015. He was a valued member of the Center for Astrophysics and a strong supporter of SPT science - he will be greatly missed by all of us.
M.\ M.\ acknowledges support for X-ray analysis by NASA through Chandra Award Numbers 13800883 and 16800690 issued by the Chandra X-ray Observatory Center, which is operated by the Smithsonian Astrophysical Observatory for and on behalf of NASA. 
The South Pole Telescope program is supported by the National Science Foundation through grants ANT-0638937 and PLR-1248097. 
Argonne National Laboratory's work was supported under U.S. Department of Energy contract DE-AC02-06CH11357.
This work was partially completed at Fermilab, operated by Fermi Research Alliance, LLC under Contract No. De-AC02-07CH11359 with the United States Department of Energy.
Partial support is also provided by the NSF Physics Frontier Center grant PHY-0114422 to the Kavli Institute of Cosmological Physics at the University of Chicago, the Kavli Foundation, and the Gordon and Betty Moore Foundation. 
DN is supported in part by NSF grant AST-1412768, NASA Chandra Theory grant GO213004B, and by the facilities and staff of the Yale Center for Research Computing.
DR is supported by a NASA Postdoctoral Program Senior Fellowship at the NASA Ames Research Center, administered by the Universities Space Research Association under contract with NASA.


\end{document}